\documentclass[a4paper]{jpconf}
\usepackage{graphicx}
\usepackage{amssymb,amsmath}
\pdfoutput=1
\begin{document}
\title{Conditional control of quantum beats in a cavity QED system}

\author{D G Norris, A D Cimmarusti and L A Orozco}

\address{Joint Quantum Institute, Department of Physics, University of Maryland and National Institute of Standards and Technology, College Park, MD 20742-4111, U.S.A.}

\ead{lorozco@umd.edu}

\begin{abstract}
We probe a ground-state superposition that produces a quantum beat in the intensity correlation of a two-mode cavity QED system. We mix drive with scattered light from an atomic beam traversing the cavity, and effectively measure the interference between the drive and the light from the atom. When a photon escapes the cavity, and upon detection, it triggers our feedback which modulates the drive at the same beat frequency but opposite phase for a given time window. This results in a partial interruption of the beat oscillation in the correlation function, that then returns to oscillate.
\end{abstract}

\section{Introduction}

Quantum feedback \cite{wisemanbook} and quantum control \cite{rabitz09} are important disciplines with relationships to quantum information science. The question of how to control a quantum system without disturbing it remains open in general \cite{habib02}, but the search for efficient protocols continues and the experimental realization is now using weak quantum measurements. (See for example the recent paper by Gillett {\it et al.} \cite{gillett10}).

The detection of a photon escaping a quantum system at a random time heralds the preparation of a conditional quantum state. Manipulation of these states is essential in the field of quantum feedback. The preferential probe of this conditional measurement in quantum optics is the intensity correlation function which has been used since the pioneer work of Kimble {\it et al.} on resonance fluorescence \cite{kimble77}. 

This paper presents the preliminary implementation of indirectly coupled quantum feedback in our cavity quantum electrodynamical (QED) system. It acts on the ground state coherences we recently observed  \cite{norris10}. However, it builds up on extensive literature that has looked into the evolution and control of quantum states such as Refs.~\cite{smith06,nielsen08}. This work closely follows our previous studies~\cite{smith02,smith04,reiner04a}, except our conditional state manipulation is long-lived ($\sim 5$ $\mu$s) and it consists of a ground state superposition detected through a homodyne measurement done in photon counting.  

Wiseman \cite{wiseman02} established the connection between homodyne measurements and weak measurements in cavity QED. Weak measurements reduce the problems of back action in quantum feedback \cite{gillett10, jacobs10}. Our previous work with conditional homodyne detection \cite{foster00,carmichael00,foster02} used a strong local oscillator. Recent measurements perform homodyne detection of resonance fluorescence with a weak local oscillator \cite{gerber09}. Our work is moving on that direction and we expect to improve our ability to control the quantum states with new forms of feedback.

\section{Description of the quantum beats}
Quantum beats are amongst the first phenomena to be fully accounted for by quantum mechanics~\cite{breit33}. They consist of oscillations in the radiation intensity of a group of excited atoms due to interfering emission pathways. Usually the atomic systems that exhibit quantum beats have the ``Type I'' (V system) energy level structure: two excited states and a ground state. The atoms are initially prepared in a superposition of the excited states, by for example a broadband excitation pulse. They then decay to the ground state, and beating occurs between the two decay paths at the difference between the frequencies of emission (the excited state splitting). For ``Type II'' ($\Lambda$ system) atoms the situation is reversed. We now have only one excited state and two ground states. References~\cite{breit33,chow75,herman75} show that no beat is possible for ``Type-II'' atoms in which the two orthogonal ground states are non-degenerate. However, while this is true for the mean intensity, beats may still lie in the fluctuations~\cite{forrester55}. Our recent work~\cite{norris10} shows the ground-state quantum beats observed in the conditional evolution of a cavity quantum electrodynamical (QED) system. They show dynamics lost on the average but retained in the variance. 

\begin{figure}[!h]
\begin{center}
\includegraphics[width=8cm]{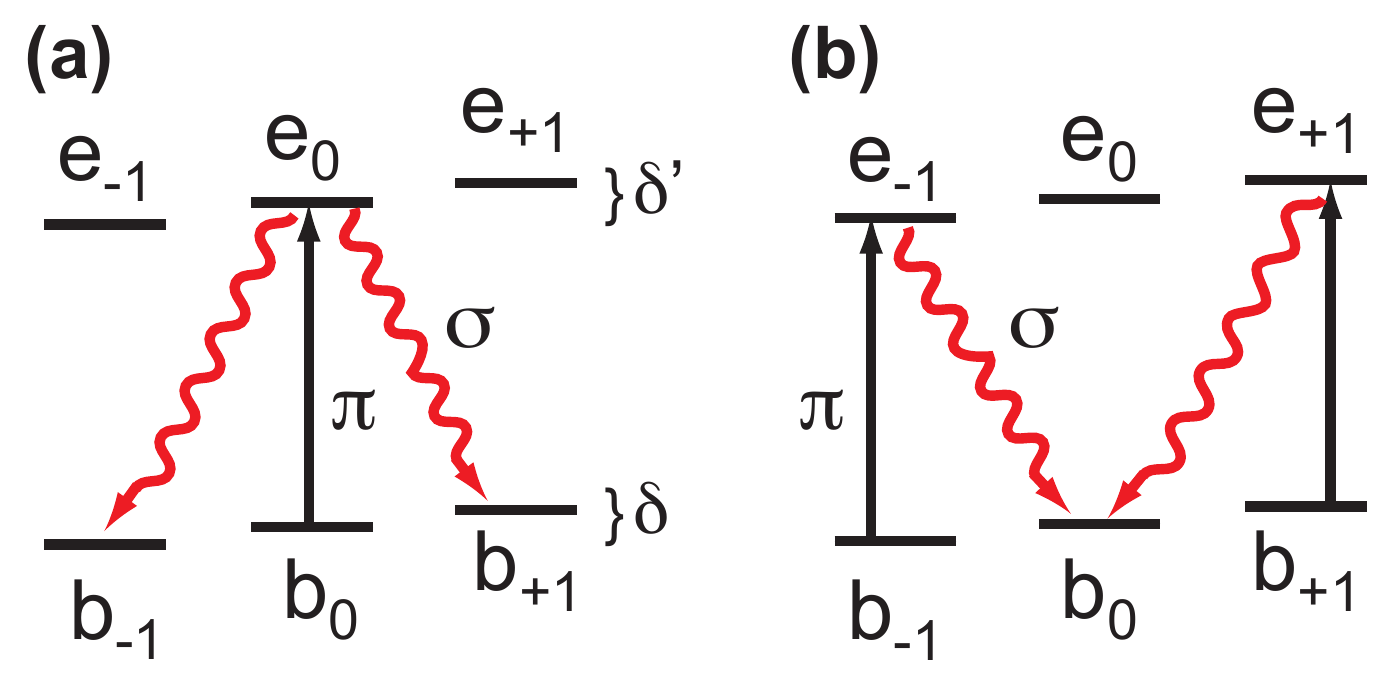}
\caption[atomic energy levels]{\label{transitions} (color online) Simplified atomic energy structure with Zeeman levels. The drive induces the $\pi$ transitions. (a) The atom decays back to the ground state, but is now in a superposition. (b) The drive reexcites the atom and it decays back to the initial state. (Figure based on Ref.\cite{barberis10}.}
\end{center}
\end{figure}

We use an optical cavity QED system in the intermediate coupling regime (\textit{i.e.} the dipole coupling constant is of the same order as the cavity and spontaneous emission decay rates) for the study of the beats. Consider a single atom with hyperfine and Zeeman structure in the ground and excited states. The atom interacts with two orthogonally polarized cavity modes, vertical (\textit{V}) and horizontal (\textit{H}). We work in the weak, continuous drive regime for the \textit{V} mode. The excitation is such that we can keep up to two photons in the undriven \textit{H} mode for intensity correlation measurements. The transitions take place between $F \rightarrow F'$ ($F \neq F'$). A weak magnetic field defines the quantization axis in the direction of \textit{V}. This mode drives $\pi$ transitions. The ground ($\delta$) and excited ($\delta'$) state Zeeman frequency shifts may be different, and we limit the discussion to the six central Zeeman sublevels of the manifold as indicated in Fig.~\ref{transitions}. The cavity decay rate is such that photon leakage can occur before reabsorption, so we neglect the latter~\cite{norris10,barberis10}.

\begin{equation}
\begin{split}
\label{condstate}
\lvert \psi_i \rangle & = \lvert b_{0} , 0 \rangle ~~~~~~~~~~~~~~~~~~~~~~~~~~~~~~~~~~~~~~~~(a)\\
\lvert \psi_i^c (t) \rangle & = \alpha e^{i \delta t} \lvert b_{-1} , 0 \rangle + \beta e^{- i \delta t} \lvert b_{+1} , 0 \rangle ~~~~~~~~~~~(b)\\
\lvert \psi_f^c (t) \rangle & = \alpha e^{i \delta t} \lvert b_{0} , 1 \rangle + \beta e^{- i \delta t} \lvert b_{0} , 1 \rangle~~~~~~~~~~~~~~~(c)
\end{split}
\end{equation}

Figure \ref{transitions} shows a simplified model for the physical origin of the quantum beats. The system starts in state $\lvert \psi_i \rangle$ (see  Eq.~\ref{condstate}a and Fig. \ref{transitions}a) with the atom at the center of the Zeeman manifold and no photons in the $H$ mode. There is a continuous excitation with $\pi$ light that at some point excites the atom to $\lvert e_0 \rangle$.  Once there, the atom will spontaneously decay to the ground state, emitting $\pi$, $\sigma^+$ or $\sigma^-$ photons. In the case of $\pi$ polarized light, the radiation adds to \textit{V}, but both $\sigma$ polarizations have components in the \textit{H} mode. Since the helicity and frequency of a $\sigma$ photon cannot be determined in the \textit{H,V} basis, the detection of a photon escaping the \textit{H} mode heralds that the atom is in a superposition state of the $m = \pm 1$ ground states: $\lvert \psi_i^c (t) \rangle$ (see  Eq.~\ref{condstate}b), that we label as the initial $(i)$ conditional $(c)$ state. This functions as the first step in our quantum eraser realization~\cite{scully82}.

The coefficients $\alpha$ and $\beta$ of the superposition depend on Clebsch-Gordan coefficients, the Zeeman ground and excited state shift difference, and the excited state linewidth~\cite{norris10}. The atom, now in the ground state, but with angular momentum perpendicular to the magnetic field, undergoes Larmor precession with a time-dependent phase  $\phi (t) = \delta t$ (see Eq.~(\ref{condstate}b), but the continuous drive \textit{V} can reexcite it (see Fig. \ref{transitions}b).
The atom then can spontaneously decay back to $\lvert b_0 \rangle$, emitting a second photon into the \textit{H} mode( Eq.~\ref{condstate}c) and leaving the system in the final ($f$) conditional ($c$) state $\lvert \psi_f^c (t) \rangle $. This second photon erases the path information present in the intermediate state and represents the second step in our quantum eraser realization.

The probability for the second emission depends on the phase $\phi (t)$ acquired since the detection of the first $H$ photon, so the quantum beats only manifest themselves in the second-order intensity correlation function, $g^{(2)}(\tau)$; they are not visible in the mean transmitted intensity. Eq.~(\ref{intensity1}) shows the calculation of the conditional intensity $\langle I_1 (t) \rangle_c $ of a second photon starting with the conditional final state $\lvert \psi_f^c (t) \rangle$. This gives the unnormalized second-order correlation function \cite{carmichael93book}.

\begin{equation}
\begin{split}
\label{intensity1}
\langle I_1 (t) \rangle_c & = \langle \psi_f^c (t) \lvert a^\dagger a \rvert \psi_f^c (t) \rangle \\
& = ( \alpha^* \beta e^{- 2 i \delta t} + \alpha \beta^* e^{2 i \delta t} + \lvert \alpha \rvert^2 + \lvert \beta \rvert^2 \\
& = 2 \lvert \alpha \rvert \lvert \beta \rvert \cos (2 \delta t + \phi_1) + \lvert \alpha \rvert^2 + \lvert \beta \rvert^2
\end{split}
\end{equation}
Here $\phi_1$ is a possible complex phase difference between $\alpha$ and $\beta$. The normalized second-order correlation function recovers the quantum beats as an oscillation at frequency $2\delta$, twice the Larmor precession frequency. 

The real experiment is more complex, as we use an atomic beam rather than a single atom. There can be many atoms in the cavity mode at any given time, with a random distribution in the Gaussian transverse profile and standing wave.  In addition, small amounts of light from the driven mode may be coupled into the orthogonal mode through cavity birefringence or optical elements. Ref. \cite{norris10} shows that even with these complications, the beats do survive, but can come from three different physical mechanisms. 

We use the work of Carmichael \textit{et al.}~\cite{carmichael78} who give the analytical form (Eq. \ref{g2func}) of the measured average second-order correlation function $\overline{g_s^{(2)}}(\tau)$ for the problem of resonance fluorescence, taking into account atomic number fluctuations in a beam. Although our system is not strictly this, since atoms in a cavity mode are not fully independent, their treatment is approximately valid under the assumption of no reabsorption of an emitted $H$ photon in our cavity:

\begin{equation}
\label{g2func}
\overline{g_s^{(2)}} (\tau) = 1 + \frac{1}{\left( 1 + \Upsilon / \bar{N} \right)^2} \left( \frac{g_A^{(2)}(\tau)}{\bar{N}} + \left| g_A^{(1)} (\tau) \right|^2 f(A) + \frac{2 \Upsilon}{\bar{N}} \text{Re} \left( g_A^{(1)}(\tau) \right) f_D(A) \right)  
\end{equation}
Here $\Upsilon$ is the background-to-signal ratio for a single atom (the background can consist of a small amount of mixed drive from the \textit{V} mode), $\bar{N}$ is the mean number of atoms in the mode, and $g_A^{(1)}(\tau)$ and $g_A^{(2)}(\tau)$ are the normalized single-atom first- and second-order correlation functions. The functions $f(A)$ and $f_D(A)$ quantify the spatial coherence within a detection area $A$ for  terms containing products of fields from different sources.  Since we collect light from a single cavity spatial and polarization mode, \textit{H}, each has a value of unity. 

The first source of beats, described previously in detail in Eqs~\ref{condstate} and \ref{intensity1}, is the $g_A^{(2)}(\tau)$ term in Eq.~(\ref{g2func}). The second source of beats is the $\left | g_A^{(1)}(\tau) \right |^2$ term, a two-atom contribution arising from interference in the time-ordering of emissions from indistinguishable atoms which we associate to a conditional intensity $\langle I_2 (t) \rangle_c$. (This is the same term that gives photon bunching in thermal light \cite{loudon83}, as observed by Hanbury-Brown and Twiss \cite{brown56}.) The last contribution comes from the $\text{Re} \left( g_A^{(1)}(\tau) \right)$ term. This is an interference between the background $\Upsilon$ and the light emitted by a single atom.  We refer to it as a homodyne beat, and it occurs at the single Larmor frequency.  We recover it in the correlation function when the background is large enough so that this term dominates. 

Equation \ref{intensity3} shows a simplified way to obtain this third term (again we consider only a single atom in the cavity, so there will be no two-atom contributions.) We calculate the conditional intensity  $\langle I'(t) \rangle_c $ (Eq.~\ref{intensity3}a,b) for a second photon starting from the atomic superposition of $\lvert \psi_f^c (t) \rangle$ that has already deposited a photon in the $H$ cavity mode and we mix a certain amount of the drive $\eta$ into the H mode such that we now have the conditional state $\lvert \psi'^c (t) \rangle = \lvert \psi_f^c (t) \rangle + \eta \lvert b_0 , 1 \rangle$. The result shows that there are two parts: We recover the single atom contribution $\langle I_1(t) \rangle_c$, but there are also other terms that we collect and label $\langle I_3(t) \rangle_c$:

\begin{equation}
\begin{split}
\label{intensity3}
\langle I'(t) \rangle_c & = \langle \psi'^c (t) \lvert a^\dagger a \rvert \psi'^c (t) \rangle ~~~~~~~~~~~~~~~~~~~~~~~~~~~~~~~~~~~~~~~~~~~~~~~~~~~~~~~~~~~(a)\\
& = \langle I_1 (t) \rangle_c + 2 \lvert \eta \rvert \lvert \alpha \rvert \cos ( \delta t + \phi_3) + 2 \lvert \eta \rvert \lvert \beta \rvert \cos ( \delta t + \phi'_3) + \lvert \eta \rvert^2,~~~~~~~(b)\\
\langle I_3 (t) \rangle_c & = 2 \lvert \eta \rvert \lvert \alpha \rvert \cos ( \delta t + \phi_3) + 2 \lvert \eta \rvert \lvert \beta \rvert \cos ( \delta t + \phi'_3) + \lvert \eta \rvert^2.~~~~~~~~~~~~~~~~~~~~(c)
\end{split}
\end{equation}
Here $\phi_3$ and $\phi'_3$ are respectively the complex phase differences between $\alpha$ and $\eta$ and $\beta$ and $\eta$. All these parameters have stable values because they are in a cavity. This homodyne term is different from the others in that it can take negative values and thus cause $\overline{g_s^{(2)}}(\tau)$ to dip below unity. (See in particular Fig. 4e in Ref.~\cite{norris10} and the relevant discussion in the text.)

The crux of our feedback protocol lies in the manipulation of this homodyne beat. We mix light from the \textit{V} mode with  light from the \textit{H} mode on the cavity output. The increasing fraction of \textit{V} mode light causes the homodyne beat term to dominate in $\overline{g_s^{(2)}}(\tau)$.  While the feedback should in principle work equally well when the other terms dominate, the larger visibility of the homodyne term makes it easier to work with experimentally.
%
\section{Experimental setup}
\begin{figure}[!h]
\begin{center}
\includegraphics[width=10cm]{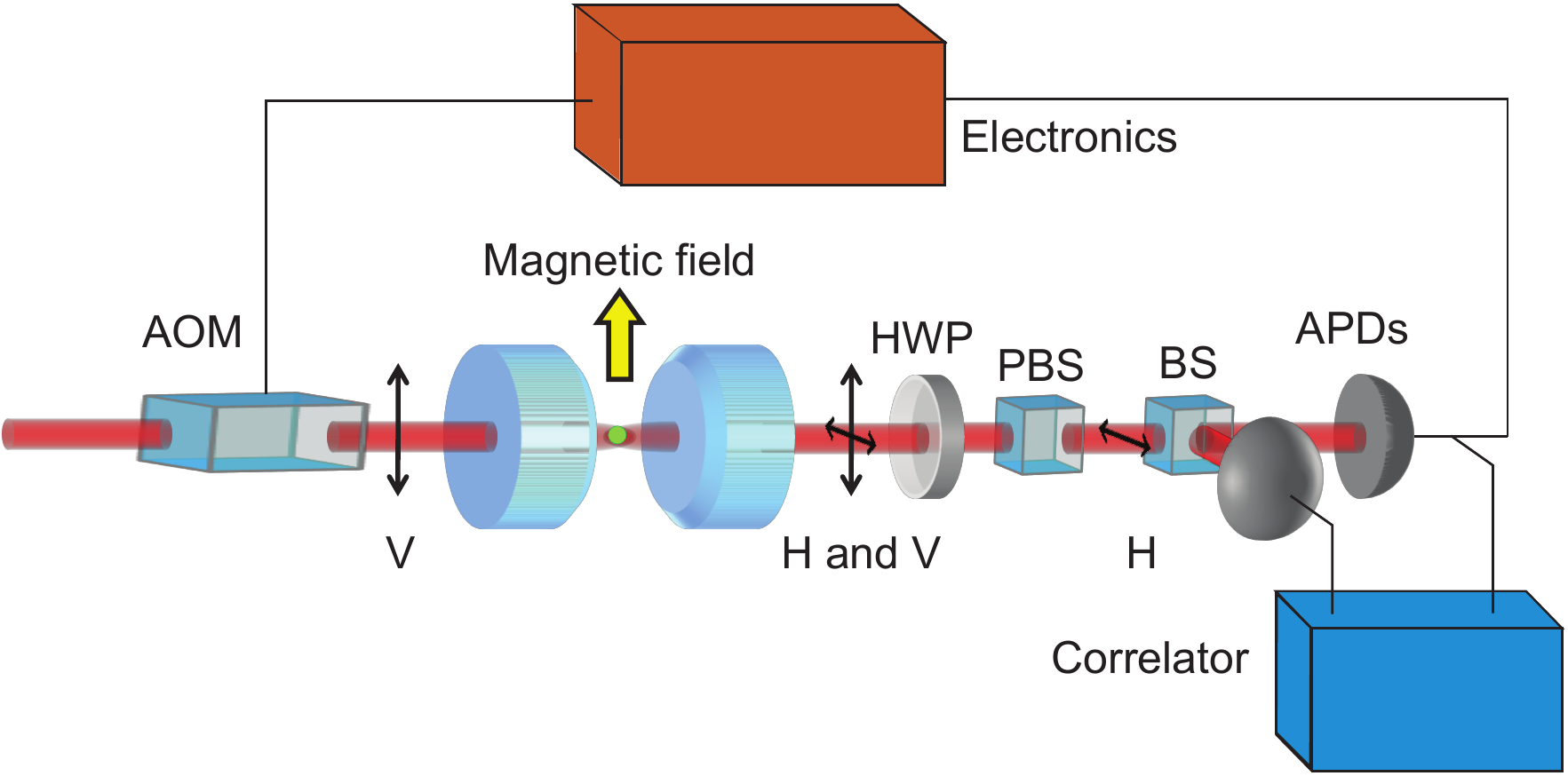}
\caption[apparatus]{\label{apparatus}(color online) Schematic of the apparatus. HWP: Half-Wave Plate, APD: Avalanche Photo-Diode, PBS: Polarizing Beam Splitter, BS: Beam Splitter, AOM: Acousto-Optic Modulator}
\end{center}
\end{figure}

Figure \ref{apparatus} shows the main features of the experiment. A 780 nm linearly polarized laser beam drives the TEM$_{00}$ $V$ mode of a Fabry-Perot optical cavity in resonance with the $D_2$ line of $^{85}$Rb. A cold beam of atoms goes through the cavity at near perpendicular incidence with respect to the mode. As the lifetime of the excited state of Rb is only 26 ns, the atoms undergo multiple excitations during the transit time of $\sim 5$ $\mu$s. The mirrors sit 2.2 mm apart, forming a 11,000 finesse cavity with decay rate $\kappa / 2 \pi = 3.2 \times 10^6~$s$^{-1}$ comparable to the atomic decay rate $\gamma / 2 \pi = 6 \times 10^6~$s$^{-1}$ and dipole coupling constant $g / 2 \pi = 1.5$ MHz, for the transition $5S_{1/2}~(F,m) = (3,0) \rightarrow 5P_{3/2}~(F',m') = (4,0)$. For a more detailed description of the apparatus, see Ref.~\cite{norris09a}. 

Atoms enter the cavity optically pumped to $5S_{1/2}~F=3, m=0$ which corresponds to our $\lvert b_0 \rangle$. A Glan-Thompson polarizer and zero-order half-wave plate (HWP) placed before the cavity linearly polarize the drive with a very good extinction ratio that can reach better than $5 \times 10^{-5}$. After the cavity another HWP aligns the output polarization to a Wollaston polarizing beam splitter (PBS) to separate the \textit{H} and \textit{V} modes. The \textit{H} light passes through a regular beam splitter (BS) which divides the light between two avalanche photodiodes (APD). Both detector outputs then go to a correlator card (Becker and Hickl DPC-230) which records a continuous stream of detection times with a resolution of 164 ps.


The rest of the components in Fig.~\ref{apparatus} enable the feedback protocol. The ``electronics box'' represents the following: The pulse from the `start' APD (designated arbitrarily) is split into two and passed through a Lecroy 688AL level adaptor to produce a clean TTL pulse.  This triggers an HP 33120 signal generator whose output controls the amplitude modulation port of an Isomet D323B radio frequency driver box.  The driver connects to an 80 MHz Crystal Technology 3080-122 acousto-optical modulator (AOM), whose first-order diffracted beam drives the cavity. In this way, the intensity of the drive can be modulated conditionally, based on the trigger from the `start' APD.
\section{Preliminary results}
\begin{figure}[!h]
\begin{center}
\includegraphics[width=10cm]{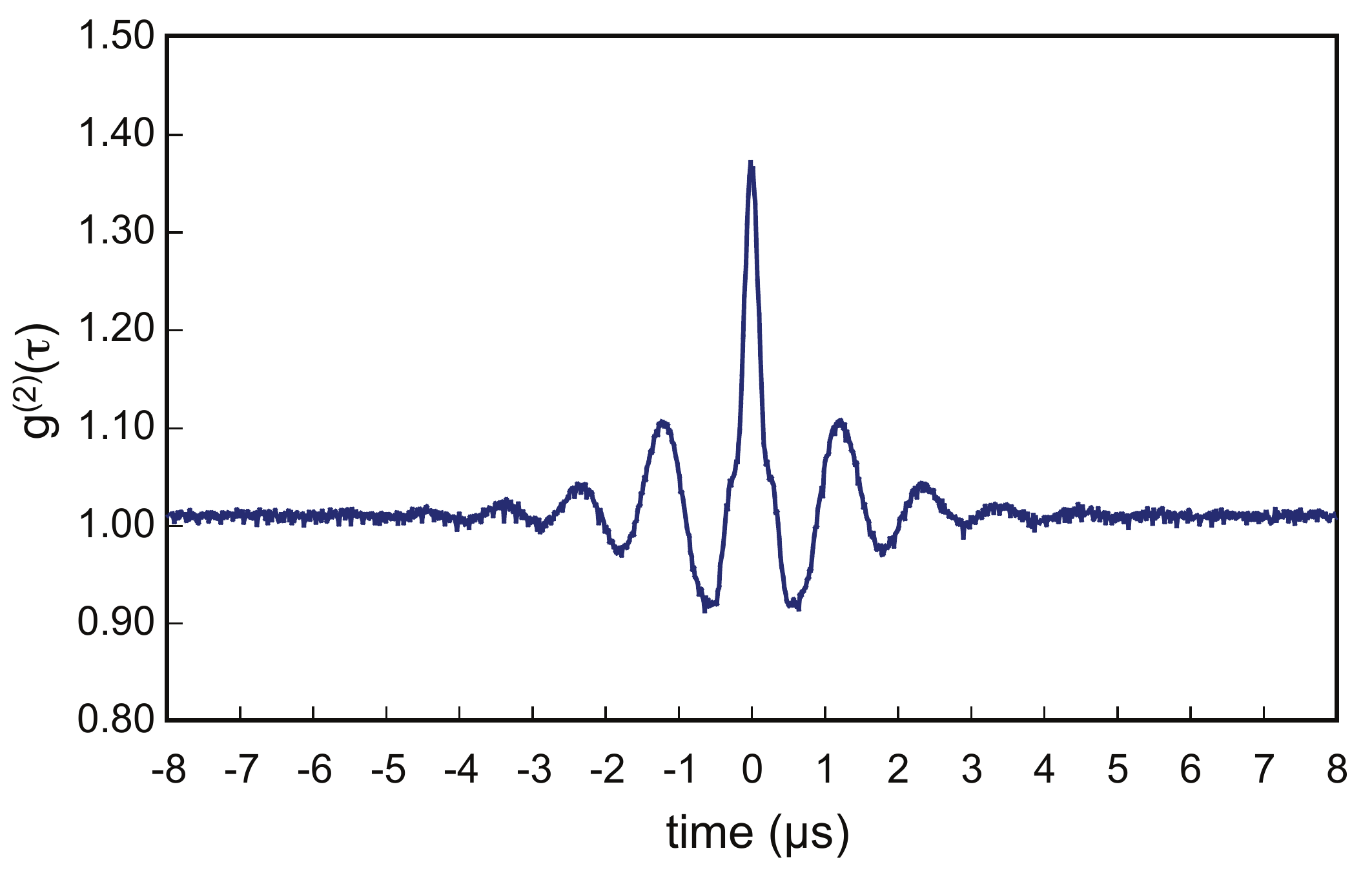}
\caption[${g^{(2)}}(\tau)$ correlation function]{\label{g2}(color online) $g^{(2)}(\tau)$ exhibiting a quantum beat oscillation with $f = 860$ KHz, corresponding to a magnetic field strength of 1.8 G}
\end{center}
\end{figure}
We measure the intensity correlation function  ${g^{(2)}}(\tau)$ from our cavity in a regime where the homodyne quantum beat term dominates, which we achieve by changing the angle on the HWP after the cavity by approximately 2 degrees away from maximum drive extinction, which increases the value of $\Upsilon$. The effective number of maximally coupled atoms in the mode is approximately 2. Fig.~\ref{g2} shows our normalized second-order correlation function due primarily to the beating against the drive; this is apparent because it dips below one. 

The basic idea for control is simple. We rely on conditional measurements to set the initial phase of the quantum beat. Since the intensity of the detected light is proportional to the drive intensity (from both the atomic spontaneous emission and the driven mode response), we can modulate the drive at the same frequency as the conditional output signal but with opposite phase. This way the beat will cancel as long as the modulation amplitude is chosen correctly.

\begin{figure}[h]
\begin{center}
\includegraphics[width=10cm]{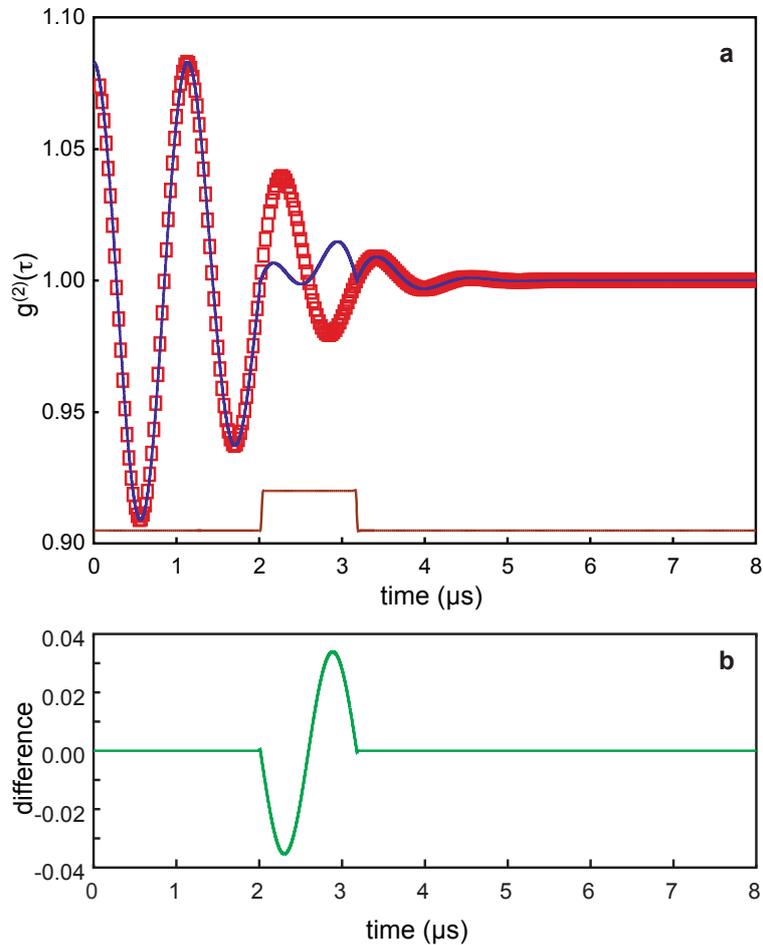}
\caption[$g^{(2)}(\tau)$ with feedback]{\label{calculated}(color online) Calculated $g^{(2)}(\tau)$ signal from the feedback model with parameters extracted from the experiment. {\bf (a)} The red squares are model without feedback. The blue trace is model calculation exhibiting the effects of our feedback. The brown trace at the bottom identifies the time window where we apply the feedback. {\bf (b)} Shows in green the difference between the red squares (no feedback) and the blue line (with feedback). }
\end{center}
\end{figure}

We are able to model the signal (after 0.5 $\mu$s) with a simple function that contains an oscillation ($\cos{\Omega t}$) at frequency $\Omega/2\pi$=860 kHz, Gaussian damping ($\exp{-(t^2/\sigma^2)}$) with $\sigma=1.8~\mu$s, and amplitude and time offsets; the intent is to capture the basic physics, not to fit the exact form. The oscillation corresponds to the Larmor frequency and the characteristic time of the Gaussian reflects the transit time of the atoms through the Gaussian transverse profile of the mode. The sharp peak at the origin is a multiatom contribution (see Fig. 4d in Ref.~\cite{norris10}) that we are not taking into account. We obtain the numbers for the model by looking at the fast Fourier transform (FFT) of the data in Fig. \ref{g2} as well as at the long term ($\approx$ 8 $\mu$s) value of the background. The width of the resonance in the FFT fits well to a Gaussian, but there is an asymmetry on the characteristic width; we average the two numbers and use that for the model. There are other frequencies visible on the FFT, coming from the standing wave modulation of the dipole coupling constant and from the harmonics of the Larmor frequency; we ignore these in the model.

Figure \ref{calculated}a illustrates the usual signal (red squares) and the signal with the feedback protocol (continuous blue line) based in the model of the signal that we just presented. It is clear that there is a modification of the response during the time that the pulse is applied, but the cancellation is not perfect. The difference Fig. \ref{calculated}b between the trace with feedback and that without recovers the applied modulation to the input drive.

A photon ``click'' in the `start' detector triggers the signal generator, which outputs a sinusoidal voltage pulse whose amplitude-to-offset ratio is 8.5\%, in a voltage region where the AOM and driver amplitude response is linear.  The delay in the application of the modulation to the drive has an intrinsic ($\sim 1.5$ $\mu$s) contribution from the signal generator, and a variable part which we use to adjust the phase to match that  of the quantum beats.

\begin{figure}[h]
\begin{center}
\includegraphics[width=10cm]{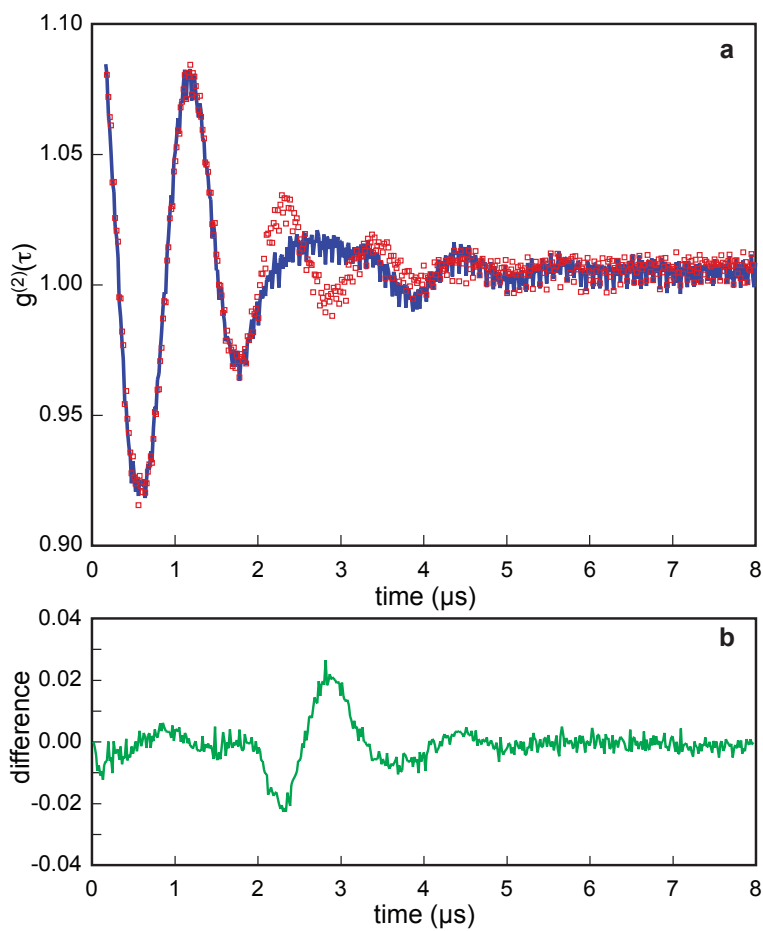}
\caption[with feedback]{\label{feedback}(color online) Experimental measurements of {\bf (a)} $g^{(2)}(\tau)$. The red squares is the negative-$\tau$ portion reflected back across the vertical axis of the data. The blue trace is the positive-$\tau$ portion, exhibiting the effects of our feedback. {\bf (b)} shows in green the difference between the red squares (no feedback) and blue traces (with feedback).}
\end{center}
\end{figure}

The feedback pulse lasts for one period of the quantum beat oscillation ($\sim 1.2$ $\mu$s), after which the beat returns with the same phase as before. We obtain a partial attenuation of the oscillations (See blue line in Fig.~\ref{feedback}a), owing primarily to the mismatch between the shape of the applied pulse and the measured $g^{(2)}(\tau)$.  Performance can be improved with use of a programmable pulse generator that matches more carefully the shape of the decaying exponential.  In addition,  trigger events missed due to signal generator dead time decrease the effects of the feedback.   


%
\section{Future work}
We wish to fully and deterministically manipulate the quantum beats exhibited by $g^{(2)}(\tau)$, in the sense of controlling the amplitude, phase, or frequency of the atoms in the ground state. 
One possibility is to use laser pulses from the cavity side to conditionally transport half of the ground-state superposition to a different state. This excitation would interrupt the coherent evolution, with the possibility of bringing it back with deterministic phase by use of coherent Raman transitions. Reference~\cite{barberis10} shows applications in quantum error correction.

Preliminary attempts with a laser perpendicular to the cavity mode have produced large mechanical effects in the atoms, pushing them out of the cavity. Modifications to the apparatus should allow retroflection of the beam to cancel such effects, as well as the application of arbitrary polarization states.
\section{Conclusions}

Our cavity QED system exhibits a long-lived homodyne quantum beat which has great potential for studies in the feedback and control of ground-state coherences. A simple feedback mechanism that modulates the drive shows moderate control of the conditional quantum beats in the intensity of the photon correlations. More elaborate techniques will further probe the nature of quantum feedback in our system.
\ack
This work was supported by the National Science Foundation (NSF). We thank Pablo Barberis-Blostein and Howard Carmichael for their stimulating discussions and continued theoretical support.
\section*{References}
%

\begin{thebibliography}{10}
\expandafter\ifx\csname url\endcsname\relax
  \def\url#1{{\tt #1}}\fi
\expandafter\ifx\csname urlprefix\endcsname\relax\def\urlprefix{URL }\fi
\providecommand{\eprint}[2][]{\url{#2}}

\bibitem{wisemanbook}
Wiseman H~M and Milburn G~J 2009 {\em Quantum Measurement and Control\/}
  (Cambridge: Cambridge University Press)

\bibitem{rabitz09}
Rabitz H 2009 {\em New Journal of Physics\/} {\bf 11} 105030
  \urlprefix\url{http://stacks.iop.org/1367-2630/11/i=10/a=105030}

\bibitem{habib02}
Habib S, Jacobs K and Mabuchi H 2002 {\em Los Alamos Science\/} {\bf 27} 126

\bibitem{gillett10}
Gillett G~G, Dalton R~B, Lanyon B~P, Almeida M~P, Barbieri M, Pryde G~J,
  O'Brien J~L, Resch K~J, Bartlett S~D and White A~G 2010 {\em Phys. Rev.
  Lett.\/} {\bf 104} 080503

\bibitem{kimble77}
Kimble H~J, Dagenais M and Mandel L 1977 {\em Phys. Rev. Lett.\/} {\bf 39} 691

\bibitem{norris10}
Norris D~G, Orozco L~A, Barberis-Blostein P and Carmichael H~J 2010 {\em Phys.
  Rev. Lett.\/} {\bf {105}} {123602}

\bibitem{smith06}
Smith G~A, Silberfarb A, Deutsch I~H and Jessen P~S 2006 {\em Phys. Rev.
  Lett.\/} {\bf 97} 180403

\bibitem{nielsen08}
Nielsen A~E~B and M\o{}lmer K 2008 {\em Phys. Rev. A\/} {\bf 77} 063811

\bibitem{smith02}
Smith W~P, Reiner J~E, Orozco L~A, Kuhr S and Wiseman H~M 2002 {\em Phys. Rev.
  Lett.\/} {\bf 89} 133601

\bibitem{smith04}
Smith W~P and Orozco L~A 2004 {\em J. Opt. B: Quantum Semiclass. Opt.\/} {\bf
  6} 135

\bibitem{reiner04a}
Reiner J~E, Smith W~P, Orozco L~A, Wiseman H~M and Gambetta J 2004 {\em Phys.
  Rev. A\/} {\bf 70} 0238119

\bibitem{wiseman02}
Wiseman H~M 2002 {\em Phys. Rev. A\/} {\bf 65} 032111

\bibitem{jacobs10}
Jacobs K 2010 {\em New Journal of Physics\/} {\bf 12} 043005
  \urlprefix\url{http://stacks.iop.org/1367-2630/12/i=4/a=043005}

\bibitem{foster00}
Foster G~T, Orozco L~A, Castro-Beltran H~M and Carmichael H~J 2000 {\em Phys.
  Rev. Lett.\/} {\bf 85} 3149

\bibitem{carmichael00}
Carmichael H~J, Castro-Beltran H~M, Foster G~T and Orozco L~A 2000 {\em Phys.
  Rev. Lett.\/} {\bf 85} 1855

\bibitem{foster02}
Foster G~T, Smith W~P, Reiner J~E and Orozco L~A 2002 {\em Phys. Rev. A\/} {\bf
  66} 033807

\bibitem{gerber09}
Gerber S, Rotter D, Slodi\ifmmode~\check{c}\else \v{c}\fi{}ka L, Eschner J,
  Carmichael H~J and Blatt R 2009 {\em Phys. Rev. Lett.\/} {\bf 102} 183601

\bibitem{breit33}
Breit G 1933 {\em Rev. Mod. Phys.\/} {\bf 5} 91

\bibitem{chow75}
Chow W~W, Scully M~O and {Stoner~Jr} J~O 1975 {\em Phys. Rev. A\/} {\bf 11}
  1380

\bibitem{herman75}
Herman R~M, Grotch H, Kornblith R and Eberly J~H 1975 {\em Phys. Rev. A\/} {\bf
  11} 1389--1396

\bibitem{forrester55}
Forrester A~T, Gudmundsen R~A and Johnson P~O 1955 {\em Phys. Rev.\/} {\bf 99}
  1691

\bibitem{barberis10}
Barberis-Blostein P, Norris D~G, Orozco L~A and Carmichael H~J 2010 {\em New J.
  Phys.\/} {\bf 12} 023002

\bibitem{scully82}
Scully M~O and Dr{\"u}hl K 1982 {\em Phys. Rev. A\/} {\bf 25} 2208

\bibitem{carmichael93book}
Carmichael H~J 1993 {\em An Open Systems Approach to Quantum Optics, Lecture
  Notes in Physics\/} vol~18 (Berlin: Springer-Verlag)

\bibitem{carmichael78}
Carmichael H~J, Drummond P, Meystre P and Walls D~F 1978 {\em J. Phys. A: Math.
  Gen.\/} {\bf 11} L121--L126

\bibitem{loudon83}
Loudon R 1983 {\em The Quantum Theory of Light\/} 2nd ed (New York: Oxford
  University Press)

\bibitem{brown56}
Brown R~H and Twiss R~Q 1956 {\em Nature\/} {\bf 177} 27

\bibitem{norris09a}
Norris D~G, Cahoon E~J and Orozco L~A 2009 {\em Phys. Rev. A\/} {\bf 80} 043830

\end{thebibliography}
\providecommand{\newblock}{}

\end{document}